\documentclass[pss]{wiley2sp} 
\usepackage{amsmath}

\begin{document}

\title{Synthesis of nitrogen doped single wall carbon nanotubes with caffeine.}

\titlerunning{Short title }

\author{%
  Filippo Fedi\textsuperscript{\Ast,\textsf{\bfseries 1}},
  Oleg Domanov\textsuperscript{\textsf{\bfseries 1}},
Paola Ayala\textsuperscript{\textsf{\bfseries 1}}
  Thomas Pichler\textsuperscript{\textsf{\bfseries 1}}.}

\authorrunning{First author et al.}

\mail{e-mail
  \textsf{filippo.fedi@univie.ac.at}, Phone: +43-1-4277-72640}

\institute{%
  \textsuperscript{1}\,Electronic Properties of Materials, University of Vienna - Faculty of Physics, Vienna\\
}
\received{XXXX, revised XXXX, accepted XXXX} 
\published{XXXX} 


\abstract{%
%
%
%
\abstcol{
Nitrogen doped single wall carbon nanotubes have many functional benefits. Doping opens the possibility to control the electronic energy levels, surface energy, surface reactivity and charge carrier density. The additional electron in the outer shell changes the electronic properties of the nanotubes when introduced into the carbon lattice. Here we present the latest findings in the in-situ doping during synthesis of single wall carbon nanotubes using caffeine as a precursor of both carbon and nitrogen. A special furnace with two heating elements allowed us to sublimate and decompose the solid precursor. Caffeine allowed us to reach a high doping percentage with high quality nanotubes directly in a one-step synthesis procedure.

   }}

%
%
\titlefigure[width=7cm,height=4.5cm]{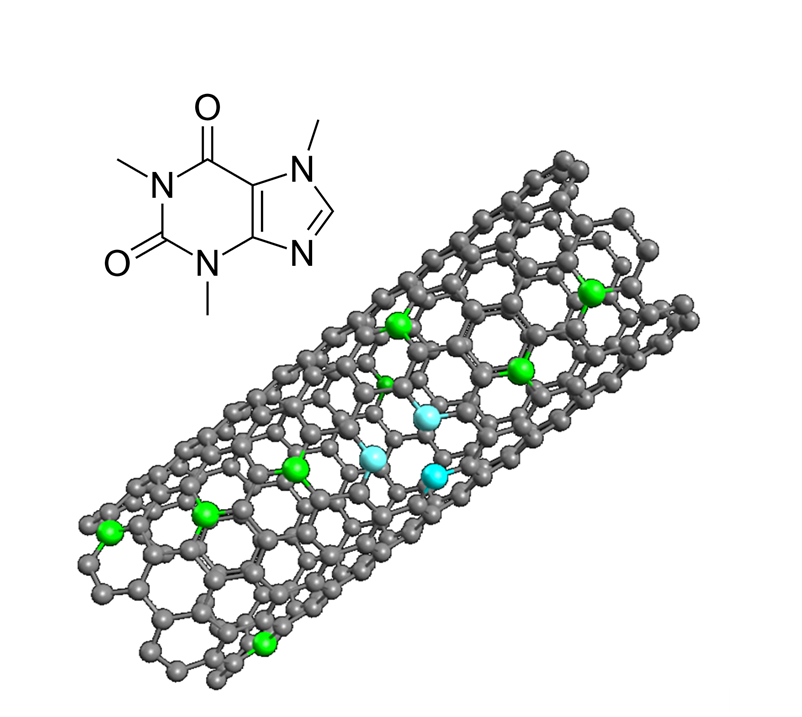}

\maketitle   

\section{Introduction}

A precise control of the structure and properties is essential for gaining on the exceptional features of carbon nanotubes (CNTs) for real-life applications \cite{harris2009,baughman2002,de2013}. CNTs in their pristine form are materials with negligible chemical reactivity and with low surface energy, insufficient to satisfy diverse applications \cite{hirsch2002,zhao2013}. A smart idea to regulate their properties is using different functionalization methods \cite{ayala2010b,sun2002,maiti2014}. In order to achieve high control of the material properties, different approaches have been proposed by different groups \cite{deng2016}.  A fascinating type of doping CNTs is by substitutionally, introducing heteroatoms into the graphitic lattice, e.g. nitrogen~\cite{terrones2002n,terrones2007,Ayala2010}. A lot of work has been done with multiwalled N-doped CNTs over the past two decades and important applications have been found~\cite{gong2009}.  Synthesizing single-walled (SW) tubes has imposed more challenges.  While some groups have achieved this by post growth treatments of pristine nanocarbons~\cite{esconjauregui2015,van2013,shrestha2017}, doping has also been obtained in situ, i.e. directly growing SWCNTs with incorporated heteroatoms from a specific precursor~\cite{ayala2007c,ayala2007tailoring,elias2010}.
Thanks to the additional electron that N contains compared to C, using this heteroatom has gained special interest over the past years~\cite{sun2014,susi2010,sharifi2012}, since some novel electronic properties have been predicted because the N-CNT would become an n-type doped material \cite{czerw2001}.
The efficiency of the incorporation of N atoms within the CNTs lattice depends on several factors: choice of precursor, catalyst type, reaction temperature and time, gas flow rate and relative pressures~\cite{ayala2007b,ayala2007effects}. Although N-CNTs have several advantages, a synthesis method for their production that ensures high quality, in terms of D/G ratio, of N-CNTs is not yet available. In fact, the quality of the materials obtained sometimes can be very low, the reproducibility is difficult and the retention of N atoms is weak and N\textsubscript{2} gas may be trapped into the CNT~\cite{terrones1999,ghosh2010}. In addition it is quite common to use toxic and corrosive precursors that are not easy to be handled.  Tuning of the electronic properties of nitrogen doped single wall carbon nanotubes (N-SWCNTs) is very challenging and if achieved in a controlled manner, a very promising material will be available for applications such as in active elements of future nano-electronic devices, optical devices and sensors.
\newline The aim of this study is to elucidate the possibility to synthesize high quality, stable N-SWCNTs starting from caffeine, a very common, cheap, and non-toxic organic molecule that contains both N and C. The innovative precursor that we employed is a purine, heterocyclic aromatic organic compound made by a pyrimidine ring fused to an imidazole ring in addition with some other specific functional groups made of carbon, oxygen and nitrogen. Furthermore, to the best of our knowledge, employing caffeine a precursor is something that has never been done so far. In order to characterize the doping level and sample quality, Raman and X-ray photoelectron spectroscopy (XPS) have been utilized.

\section{Materials and methods}
N-SWCNTs were synthesized using catalytic chemical vapor deposition (CCVD) with a two stage furnace in high vacuum conditions. Caffeine by Sigma-Aldrich (ReagentPlus\textsuperscript{\textregistered}) was used, which is a powder. The catalyst as produced mixing 3wt.\% ammonium iron citrate (Sigma-Aldrich) with 97 wt.\% magnesium oxide (Sigma-Aldrich) in ethanol, bath-sonication for 72 hours, and drying at $70^\circ\text{C}$ for 24h as described elsewhere \cite{shi2015}. The growth temperature was set to $850^\circ\text{C}$ in a home built quartz furnace with a base pressure of 8 x 10\textsuperscript{-7} mBar while the pressure of the carrier gas (Argon) was 5 x 10\textsuperscript{-6} mBar. After the synthesis, the as-grown N-doped SWCNTs were subjected to a purification process. The obtained powder material was soaked for 24 hours in HCl (Sigma–Aldrich, ACS reagent, 37\%) to remove most of the MgO and catalyst particles. In order to extract the CNTs, the liquid phase was removed with by filtration through membrane (MF-Millipore, 0.20$\mu$). Afterward we collected a N-SWCNTs thin film for the further characterizations. Raman spectroscopy was performed with a Horiba Jobin Yvon LabRAM HR800 Raman spectrometer under ambient conditions with 633nm laser, 0.5 mW laser power and spectral resolution of  $\sim$2 cm\textsuperscript{-1}. XPS analysis was done using monochromatic AlK$\alpha$ radiation (1486.6 eV) and a hemispherical SCIENTA RS4000 photoelectron analyzer operating with a base pressure of 6x10\textsuperscript{-10} mBar with an overall spectral resolution of 0.5 eV.

\section{Results and discussion}

The goal of this study was to grow high quality N-SWCNTs using a solid precursor that contains carbon and nitrogen, expecting to increase the incorporation of N heteroatoms in the lattice of the SWCNT compared to previously reported work.  The choice to use a single precursor for both N and C has been already proved by several groups using melamine \cite{gaillard2005}, acetonitrile~\cite{glerup2003,thurakitseree2012,thurakitseree2013}, imidazole~\cite{ghosh2010} and benzylamine~\cite{ayala2007}. It is worth mentioning that all the abovementioned are ofter either in liquid or gas form, whereas caffeine is a powder and for that reason we adapted our system to a two stage one. In this work, caffeine has been chosen because is not poisonous, easy to manage, and with an easy highly scalable production. According to Ghosh et al.\cite{ghosh2010}, during pyrolysis these C and N feedstocks supply the pre-existing C–N fragments on the catalyst's surface, which consequently allows for the incorporation of the N atoms into the nanotubes matrix.

\begin{figure}[ht!]%
\includegraphics*[width=\linewidth,height=12.4cm]{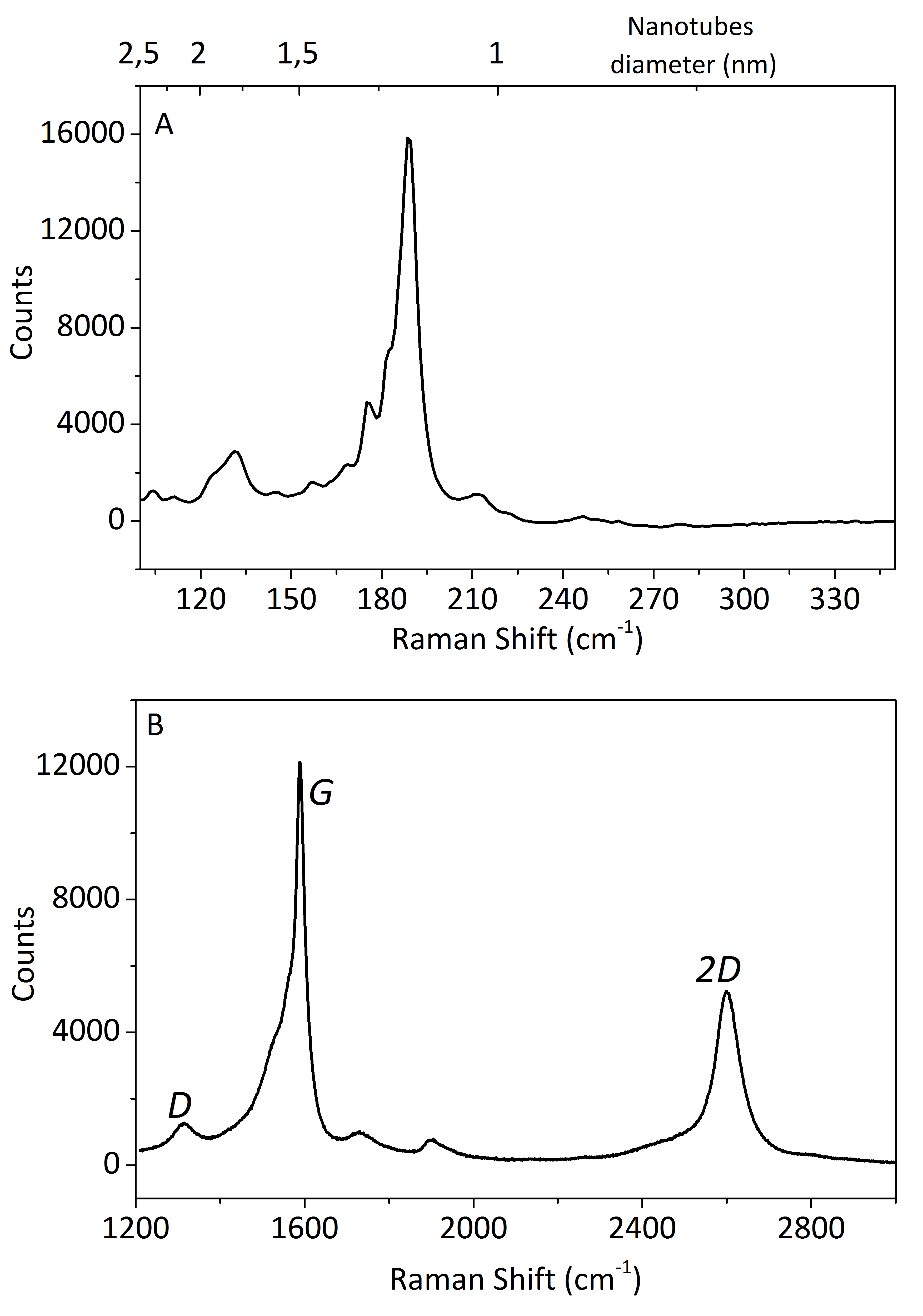}
\caption{%
A)Raman spectrum of the low-energy section of the N-SWCNT. The RBM have a narrow distribution, with the diameter between 2.1 and 1.1 nm. B) Raman spectrum of the higher-energy section of the N-SWCNT. The nanotubes show a small D-band as confirmation of their purity.
}
\label{onecolumnfigure}
\end{figure}

Our method was proved successful. The product of the synthesis were black powders similar to that of typical N-CNTs. At first, we characterized the material using Raman spectroscopy to inspect the nanotube characteristic features. Fig.~\ref{onecolumnfigure} shows in (A) the radial breathing mode (RBM) region, which is the typical fingerprint of SWCNTs~\cite{dresselhaus2005,rao1997}.  

According to Kuzmany et al \cite{kuzmany2001}  is possible to extrapolate the values of the diameters from the RBM frequencies using the formula:

 \begin{equation}
\label{eq1}
\nu[cm\textsuperscript{-1}]=\frac{234}{d[nm]}+C\textsubscript{2}
\end{equation}

A predominant peak around 190 cm\textsuperscript{-1} is observed from the high resonance of the tubes corresponding to the matching laser excitation. Nevertheless, carrying out measurements with more than one laser line have allowed us to identify a narrow distribution, with diameters between 1.1 and 2.1 nm. In Fig.~\ref{onecolumnfigure}B, the Raman spectrum of the higher-energy section embracing the D and G bands of the N-SWCNT is shown. The small D-band intensity, which is $\sim$20 times smaller than that of the G band is a
clear signature of the quality of the single-wall material and the low content of other carbonaceous species. The D peak located around $\simeq$1348 cm\textsuperscript{-1} is related to an inter-valley double resonance Raman associated to defects, vacancies in the lattice, distortions and out-of plane atoms \cite{maciel2008,campos2010}. In our material we can observe a small D-band with ratio of the area of the D band and G band (I\textsubscript{D}/I\textsubscript{G}) of only 0.04. Subsequently to the Raman characterization, we performed a wet purification, meaning soaking of the as-grown material for 24 hours in HCl.

Focusing again on the RBM region, the measurements with the 633 nm excitation wavelength, a small window of highest abundance around the mean diameter centered at 1.2 nm has been clearly identified ~\cite{kuzmany2001}. It is interesting to observe the narrow distribution of the peaks in the low-frequency bands.

\begin{figure}[h]%
\includegraphics*[width=\linewidth,height=12.4cm]{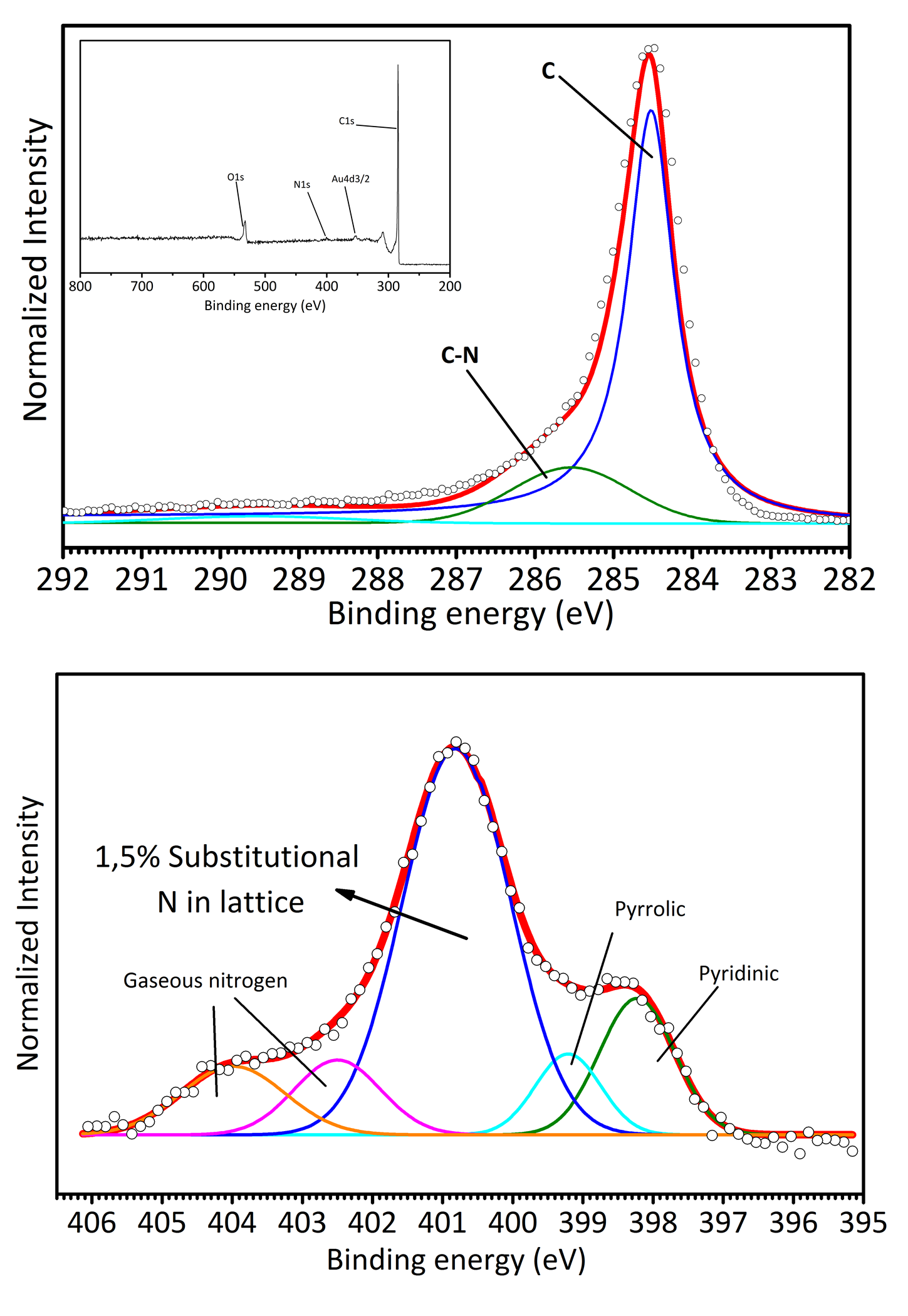}
\caption{%
Insert top figure: XPS survey of the material. The main peak around 285 is related to C, then are shown the peak of N and O.
Top figure: line shape analysis of C1s, where the peak at 284.5 eV corresponds to the C in sp\textsubscript{2} configuration, while the one at 285.6 eV is related to bonding C-N.
Lower figure: line shape analysis in the N1s region. The voigtians (from right to left) are related to the different types of nitrogen: pyridinic, pyrrolic, substituional N, and other gaseous species arising from the synthesis conditions. }
\label{xpsfigure}
\end{figure}

XPS was employed to study analytically the material chemical composition, the nitrogen content and the its bonding environment.
The survey spectrum recorded on the nanotubes sample shown in figure \ref{xpsfigure} has three main lines: the C1s line of carbon around 285 eV as the most prominent line, the N1s line of nitrogen around 400 eV and oxygen around 530 eV. The last one is related to the minimal oxygen remaining of the surface originated during the purification procedure and also the air manipulation of the sample before the measurements. This oxygen may be reduced with annealing at high temperature, but we refrained from doing this treatment to avoid causing  damages to the doping species. Important to notice is the absence of any metals, metal oxides, filter residues or other contaminations in the survey, indicating the high quality of the N-SWCNTs film. Note that the only signal that appears on the survey that could represent a foreign element corresponds to gold which is related to the sample holder of our spectrometer. Apart from that, no other signals have been detected within the experimental limit.  The line shape analysis of the C1s line is shown in the \ref{xpsfigure} on the top figure, where the main peak at 284,5 eV is related to the carbon with sp\textsuperscript{2} configuration. The broad peak at 285.6 eV is related to the C-N bonds, which is the first evidence that nitrogen is actually incorporated in the carbon lattice \cite{ruiz2014,lv2012}. One of the main goals of this work is to unambiguously confirm the effective presence of nitrogen in the nanotubes which has been successfully proved by the N1s core level measurements,  taking into account the different atomic cross sections of C and N. Considering all types of possibilities in which N bonds, we observed a total amount  of 2.56\% N. This value has been estimated considering the ratio of the areas of the peaks C and C-N in the C1s line. Furthermore have analyzed the different types of N bonding environments along the wall of the wall of the SWCNTs, which has been done taking into account previously reported work~\cite{Ayala2010,ayala2007b}. The N1s line is shown in the lower part of \ref{xpsfigure}, and it is mainly composed of four types of nitrogen-functional groups: pyridinic ($\simeq$398 eV), pyrrolic ($\simeq$399 eV), substitutional ($\simeq$400,7 eV ) and nitrogenated gaseous compounds or N\textsubscript{2} gas molecules ($\geq$ 402 eV) which are present at the nanotubes surface. Performing the line shape analysis, a high value of 1.55\% substitutional sp\textsuperscript{2} nitrogen and of 0.4\% of pyridinic nitrogen was observed. This value is above most of the other reported values for N-SWCNTs. Note that the XPS survey spectrum show the absence, within the experimental limit of 1\%, of any magnesium line, that could derive from the catalyst used. This is a confirmation of the favorable outcome of the purification process.

 \section{Conclusion}

In this work we successfully achieved the growth of high quality N-SWCNTs using a novel precursor, caffeine, which is not toxic, easy to handle, to synthesize and highly scalable production. The nanotubes have a high content of substitutional nitrogen atoms in the lattice of 1,55\%, contributing to the doping of the system, which is higher than most of the previously reported and experimentally confirmed values. This paves the way to further studies with this precursor with the scope to increase the control of the specific type of doping in order to tailor the properties of the nanotubes. The next steps will be the achievement of higher purity and the incorporation of these nanotubes in devices like e.g. gas-sensors with the scope to employ their increased reactivity.


\begin{acknowledgement}
This work was supported by the Austrian Science Fund through Project FWF P27769-N20 and by the EU project (2D-Ink FA726006).
PA would like to acknowledge the contribution of the COST Action CA15107 (MultiComp).
\end{acknowledgement}

%
\bibliographystyle{pss}
%

\providecommand{\WileyBibTextsc}{}
\let\textsc\WileyBibTextsc
\providecommand{\othercit}{}
\providecommand{\jr}[1]{#1}
\providecommand{\etal}{~et~al.}


\end{document}